\documentstyle{article}

\author{Zhen Wang\\
Physics Department, LiaoNing Normal University, DaLian, PC:116029,  P. R. China}
\title{Calculation of Planetary Precession with Quantum-corrected Newton's  Gravitation\thanks{The author would like to thank Prof. Guoying Chee for   beneficial discussion}
}
\date{April 25, 1998
}
\input tcilatex

\QQQ{Subject}{
}

\QQQ{Keywords}{
planetary precession, quantum
}

\begin{document}

\maketitle
\begin{abstract}
With consideration of quantization of space, we relate the Newton's
gravitation with the Second Law of thermodynamics. This leads to a
correction to its original form, which takes into consideration of the role
of classical measurement. Our calculation shows this corrected form of
gravitation can give explanation for planetary precession.
\end{abstract}

The most distinctive feature of quantum mechanics is the concept of
measurement. It is more reasonable and closer to reality compared with that
in classical physics. Therefore the first step that leads to a successful
combination of quantum mechanics and general relativity should be the
introduction of the role of measurement into classical physics. Such attempt
has been scarcely seen because most physicists will not give up the concept
of independent objectiveness of reality in classical physics. But as a
matter of fact we can only talk about the part of the {\em reality} we are
able to measure, which is certainly under the influence of our measurement.
It is the purpose of this paper to show with an example that this philosophy
of quantum physics may also work in classical physics.

We know in both classical and quantum physics, in reality and in philosophy,
nothing can be made up of zeros, otherwise many paradoxes like Zeno paradox
[1] would arise. Therefore it is natural and reasonable to assume that there
must be a basic measuring unit in every single measurement, which can not be
measured itself. It is the basic brick that constructs the result of our
measurement. I would like to call it {\em uncertainty quantum}, since we are
uncertain about its nature in principle. For example, in order for the
concept of length to make sense, there must be a length quantum. And time
would have no meaning if there were not a time quantum. Evidently this
quantum is characteristical of an observer. In this way, we have introduced
a subjective feature into classical physics. We shall see that with this
simple correction, we can explain with some precision the planetary
precession with Newton's' gravitation law, which formerly can not be
calculated without Einstein's theory of relativity. Such attempt may not be
meaningless when one consider all the futility in synthesizing Einstein's
relativity and quantum theory.

For any distance $L$ measured by an observer, there must be a space (or
length) quantum $q_l$ . $L$ will be an integer if measured in $q_l$ . Thus
the system actually contains the following states: the two mass points
separated by $L,\,L-q_l,\;L-2q_l,\cdots \cdots ,\,q_l\,$. When we observe
such system, we can only sense the overall gravitational effect of the
system, rather than that of single mass point. That is, for a system of two
mass points separated by a distance of $L$ , we can not {\em sense
gravitationally} any difference between these states. Therefore we can
assume a state weight for the system: the weight for each state is $1/L\;,$%
where $L$ is an integer. In the same way we know that the weight of every
state in the system composed of two mass points separated by $(L-1)\,q_l$ is 
$\frac 1{L-1}$ . From the famous Second Law of thermodynamics we know that
the system should evolve to states with larger and lager statistical
weights. That gives rise to gravitational interaction. As is usually done in
nonequalibrium statistical physics[2], we presume that the intensity of this
interaction should be proportional to the increment of the weight of the
state. That is

$$
F\propto \frac 1{L-1}-\frac 1L\;\;\;\;\;\;\;\;\;\;\;\;\;\;\;\;\;(1) 
$$
This changes to the following form when we use common measuring unit

$$
F=G(q_l)m_1m_2\frac 1{L(L-q_l)}\;\;\;(2) 
$$
where $G(q_l)$ is gravitational constant and we are unable to give the exact
theoretical expression for it now. When $q_l$ is too small compared with the
distance {\it L}, (2) changes to the familiar form of Newton's gravitation.
But we shall see that it is this small approximation that wipes out the
effect of measurement itself as well as, at least partly, the planetary
precession.

The orbital equation for Newton's gravitation is [3]%
$$
\frac{d^2u}{d\theta ^2}+u=\frac{k^2}{h^2}\;\;\;\;\;\;\;\;\;\;\;\;(3) 
$$
where $r=\frac 1u$ and $\theta $ are the polar coordinates of the planet. $%
k^2=GM\,$, $M$ is the mass of the sun, $h=\frac{2\pi ab}\tau $ , $a$ and $b$
are long and short radius of the orbital ellipse, $\tau $ is period of the
planet. When corrected gravitational formula (2) is employed, (3) changes to

$$
\frac{d^2u}{d\theta ^2}+(1-\frac{q_lk^2}{h^2})\,u=\frac{k^2}{h^2}\;\;\;(4) 
$$
in which terms of higher order of $q_l$ are ignored. It is straightforward
to get the solution of (4) :

$$
r=\frac p{1+Ap\cos x\theta }\;\;\;\;\;\;\;\;\;(5) 
$$
where $p=\frac{h^2-q_lk^2}{k^2}$\ , $x=\sqrt{1-\frac{q_lk^2}{h^2}}$\ , $A$
is an integral constant. The orbit described by (5) is also a periodical
function with its period to be $\frac{2\pi }x$ . The perihelion is at $%
\theta =\frac{2n\pi }x,\;n\,\;\,$is any integral. Therefore its centurial
precession is

$$
\triangle \theta =2\pi (\frac 1x-1)\cdot \frac{100}\tau
\;\;\;\;\;\;\;\;\;\;(6) 
$$

The key point in this calculation is how to decide the space quantum $q_l$ .
We take the observation error [4] to be the uncertainty quantum. Judging
from the decimal figure of the observed data, we see that the error for the
precession is $\delta =0.01"\sim 0.05"$. This is not the systematic
observation error, which may be at least partly responsible for the
deviation of our calculation. So we get the space quantum in this way: $%
q_l=\delta b$ , where $b$ is the short radius of the planetary orbit. From
the result of our calculation we can see the subjectively corrected Newton's
gravitational law can give quite satisfactory result. Though it seems no
better than general relativity, it takes into consideration of the function
of measurement for the first time in Newton's dynamics.

\begin{table}
 \begin{tabular}{|c|c|c|c|c|c|} \hline
  planet & observation & relativity & $\delta=0.01$ & $\delta=0.05$ 
         & $\delta=0.0398$ \\ \hline 
  Mercury & $43.11\pm 0.45$ & 43.03  & 10.8 & 54.1 & 43.08 \\ \hline
  Venus & $8.4\pm 4.8$ & 8.63 & 5.1 & 25.4 & 20.18 \\ \hline
  Earth & $5.0\pm 1.2$ & 3.84 & 3.1 & 15.4 & 12.30 \\ \hline
 \end{tabular}
 \caption{Calculation of planetary precession(in sec.)}
\end{table}

There exists some deviation from the observation because there may be some
minor and delicate difference between the exact meaning of observation error
and that of the error in our space quantum correction. A mediate value $%
\delta =0.398"$ gives good calculation for Mercury, but not so good for the
others. This reasonably indicates that different space quanta underlie
different measurements. Therefore we may anticipate that the amount of
precession should decrease with the increasing precision of the {\em methods}
of measurements in a certain range. This can be interpreted as due to the
structural change in spacetime, or rather to the change of the uncertainty
quantum involved in the observation. Such concept of measurement is
consistent with quantum mechanics. The uncertainty principle in quantum
physics actually tells us that the result of measurement depends on the
ignorance of the variables that are not commutable with the one being
measured. We hope that effort in unifying the concepts of measurement in
classical and quantum physics may pave the way to the unification of the two
realms.

{\bf REFERENCES}

1. R. Penrose, {\it The Emperor's New Mind, \/}(Oxford University Press,
1989)

2. I. Prigogin, {\it From Being to Becoming} (W. H. Freeman and Company,
1980)

3. YanBo Zhou, {\it Theoretical Dynamics} (Jiang Su Sience and Technology
Press, 1961, in Chinese)

4. Kenneth R. Lang, {\it Astrophysical Formulae} (Springer-Verlag, 1974)

\end{document}